\documentstyle{article}

\input pz.sty
\input epsf.sty

\begin{document}

\PZhead{3}{26}{2006}{27 February}

\PZtitletl{Photometric observations of Supernovae}{2000E, 2001B, 2001V, 
and 2001X}

\PZauth{D. Yu. Tsvetkov}
\PZinst{Sternberg Astronomical Institute, University Ave.13,
119992 Moscow, Russia; e-mail: tsvetkov@sai.msu.su}

\begin{abstract}
CCD $BVRI$ photometry is presented for two type Ia
supernovae 2000E and 2001V, for SN Ib 2001B and SN II-P 2001X.
The parameters of light curves and absolute magnitudes at maximum
light are estimated. It is shown that all four supernovae are typical
for their classes considering the shape of their light curves
and maximum luminosity.  
\end{abstract}

\begintext

\PZsubtitle{Introduction}

Continuing the long-term program of supernova (SN) observations at
Sternberg Astronomical Institute, we carried out photometry
of bright SNe 2000E, 2001B, 2001V and 2001X.
\medskip

SN 2000E was discovered by Valentini et al. (2000) at 
magnitude $V$=14.3 on CCD images
obtained with the Teramo 0.72-m TNT telescope on January 26.73 UT. 
SN was located at $\alpha = 20\hr37\mm13\sec.8,
\delta=+66\deg05\arcm50\arcs.2$ (equinox 2000.0), which is 
$6\arcs.3$ west and $26\arcs.7$ south of center of the Sbc galaxy
NGC 6951.
Turatto et al. (2000) reported that spectra obtained with 1.8-m
telescope at Cima Ekar on January 27.83 UT indicated 
that SN 2000E was type Ia event, few days before maximum light. 
CCD photometry of this SN was reported by Vinko et al. (2001)
and Valentini et al. (2003).
\medskip

The discovery of SN 2001B was reported by Xu and Qiu (2001)
on behalf of Beijing Astronomical Observatory SN survey on 
January 3.61 UT at magnitude 15.5, estimated from unfiltered CCD
image. SN was located at $\alpha =4\hr57\mm19\sec.24, 
\delta =+78\deg 11\arcm 16\arcs.5$ (equinox 2000.0),
which is $5\arcs.6$ west and $8\arcs.9$ south of the nucleus of 
Sc galaxy IC 391. Chornock and Filippenko (2001) reported that spectra
of SN 2001B, obtained on January
23 with the Shane 3-m reflector at Lick Observatory, revealed
that the object was probably a type Ib SN, roughly one week
past maximum brightness.
\medskip  

SN 2001V was discovered by Berlind on February 19.38 UT with the 
F. L. Whipple Observatory 1.5-m telescope (Jha et al., 2001).
The following precise position was determined for SN 2001V:  
$\alpha= 11\hr57\mm24\sec.93, \delta =+25\deg12\arcm09\arcs.0$ 
(equinox 2000.0),
which is $52\arcs$ east and $28\arcs$ north of the nucleus of 
edge-on Sb galaxy NGC 3987.
The spectrum of SN exhibited a blue continuum with
broad features, identifying this as a type-Ia SN well before
maximum light. Photometry of SN 2001V was later reported 
by Vinko et al. (2003).
\medskip

The discovery of SN 2001X 
was reported by Li et al. (2001)
on behalf of Beijing Astronomical
Observatory SN survey. The object was found
(magnitude about 17.0) on an unfiltered image
taken with the BAO 0.6-m telescope on February 27.8 UT. SN 2001X was
located at $\alpha=15\hr21\mm55\sec.45, \delta = +5\deg03\arcm42\arcs.1$ 
(equinox
2000.0), which is $15\arcs.5$ west and $32\arcs.4$ south of the nucleus of 
Sbc galaxy NGC 5921.
Chornock et al. (2001) reported that a spectrum of SN 2001X, obtained with 
the Lick Observatory Shane 3-m telescope
on March 3.5 UT under poor conditions, showed a very blue continuum
with strong H Balmer lines, indicating that it was a type-II
SN before maximum light. 
\bigskip
\bigskip

\PZsubtitle{Observations and reductions}

The observations were carried out at 60-cm reflector of 
Crimean Observatory of Sternberg Astronomical Institute (C60)
using SBIG ST-7 CCD camera and with 30-cm refractor (M30)
and 70-cm reflector (M70), both with SBIG ST-6 camera, in Moscow.
All reductions and photometry were made using IRAF.\PZfm 
\PZfoot{IRAF is distributed by the National Optical Astronomy Observatory,
which is operated by AURA under cooperative agreement with the
National Science Foundation}

The color terms for transformation of instrumental magnitudes $bvri$
to standard $BVR_cI_c$ were determined for different
observing seasons and telescope-filter-detector
combinations using observations of standards in M~67 
(Chevalier and Ilovaisky, 1991) and in NGC~7790 (Stetson, 2000).
The equations $b=B+K_b(B-V)+C_b; v=V+K_v(B-V)+C_v; 
r=R+K_r(V-R)+C_r; i=I+K_i(R-I)+C_i$ were solved for color terms,
which are listed in Table 1.
The images of SNe with comparison stars are shown in Figs. 1-4.
The magnitudes of comparison stars were determined on 
photometric nights, when we observed standards from Landolt (1992) and
standard regions in clusters M~67, NGC~7790, and M~92  
({\tt http://cadcwww.hia.nrc.ca/cadcbin/wdb/astrocat/stetson/query/
NGC6341}).\newline
Some magnitude estimates for comparison stars were obtained during 
observations of SNe, but they were verified later, in 2002-2004,
using observations on C60 and M70 equipped with CCD cameras
Apogee AP-7p and Ap-47p; Roper Scientific VersArray1300B.
Final values for magnitudes of comparison stars were determined by averaging
data from 5-8 nights; they are presented in Table 2,
where the designations of stars consist of the galaxy name and the 
star number on the chart. 

Photometric measurements of SNe were made relative to 
comparison stars using PSF-fitting with IRAF DAOPHOT package.
The background of host galaxies around SNe was quite smooth in most cases,
which was not surprising,
taking into account high focal ratio of our telescopes. 
Subtraction of images of host galaxies, obtained when SNe
were no longer detectable, was applied in some cases for SNe
2001B and 2001X, and the results were found practically identical
to those obtained without subtraction.
The results of SNe observations are presented in Tables 3-6,
and the light curves are shown in Figs. 5-8.
\bigskip
\bigskip
\PZsubtitle{Results and conclusions}

{\bf SN 2000E}. The light curves are shown in Fig. 5. They 
appear typical for SN Ia, the data are in good agreement with
the template light curves of SN Ia 1991T (Lira et al., 1998),
except that the decline rate at late stage for SN 2000E is higher 
than for SN 1991T.   
Our magnitudes are in good agreement with results of
Valentini et al. (2003) and Vinko et al. (2001). But our $B$ 
magnitude of 
comparison star 1 differs significantly from the value given by
Vinko et al. (2001): $\Delta B=0.18$. 
The magnitudes of SN at maximum light and at the inflection point can
be determined:
$B_{max}=14.35$ on JD 2451577, $V_{max}=13.80$ on JD 2451579,
$R_{max}=13.50$ on JD 2451577, $I_{max}=13.55$ on JD 2451575,
$B_K=17.1$ on JD 2451607.
SN 2000E had quite slow decline past maximum: $\Delta m_{15}(B)=0.95$,
$\beta_B=0.083$ mag/day, $\beta_V=0.05$. But the rate of decline
after the inflection point was close to the mean values for SNe Ia:
$\gamma_B=0.015, \gamma_V=0.019, \gamma_R=0.027$.

SN 2000E clearly suffered significant extinction both in our Galaxy
and in the host galaxy. The Galactic extinction in the direction
of NGC 6951 is $A_B=1.57$ according to Schlegel et al. (1998),
but Burstein and Heiles (1982) gave much smaller value $A_B=0.88$.
Valentini et al. (2003) reported that interstellar NaID lines in the
spectrum of SN 2000E, which originated in the host galaxy, had
equivalent width EW(NaI)$\sim$ 0.6 \AA. According to Turatto et al.
(2003) this value of EW(NaI) may correspond to color excess
$E(B-V)$ from 0.1 to 0.3 magnitudes. So, using these data, we can 
estimate total reddening of SN 2000E to lie between 0.32 and 0.70.
Comparing the color $(B-V)$ curve of SN 2000E with the color
curves for SNe Ia with negligible extinction in host galaxies, we 
can estimate $E(B-V)\approx 0.6$. Valentini et al. (2003) have
adopted $E(B-V)=0.5$ as the most likely value, after considering 
different methods for its determination. If we take this value
for reddening and distance modulus $\mu=31.93$ from LEDA 
({\tt http://leda.univ-lyon1.fr/}),
which is computed using radial velocity of the host galaxy, corrected
for Virgocentric flow, and $H_0=70$ km s$^{-1}$Mpc$^{-1}$, then we 
obtain for absolute 
magnitude of SN 2000E $M_B=-19.6, M_V=-19.7$. This is clearly 
brighter than average value for SNe Ia and confirms the 
relation between rate of decline and absolute magnitude for 
SNe Ia (Pskovskii, 1977, Phillips, 2005).
\medskip

{\bf SN 2001B}. The light curves are shown in Fig. 6, where we
also plotted the data, reported by amateur astronomers at \newline 
{\tt http://www.astrosurf.com/snweb2/2001/01B\_/01B\_Meas.htm}.
Their results are in satisfactory agreement with our data, and 
we can assume that the peak of $R$-band light curve is 
correctly determined by observations of amateurs. Supposing 
the $V$ light curve had the same shape of the peak, we fitted it
with $V$-band light curve of SN Ib 1983N (Cappellaro et al., 1995), 
and the agreement seems quite satisfactory. We can estimate 
the magnitudes at maximum light: $V_{max}=15.0$ on
JD 2451929, $R_{max}=14.4$ on approximately the same date.
The rate of linear decline after the inflection point was 
$\sim 0.015$ mag/day both in $V$ and $R$ bands. 
It is difficult to estimate extinction in the host galaxy for
SN 2001B, because color curves of type Ib/c SNe have 
considerable scatter. But in this case we can suppose that it
was not large, because reports on spectroscopic observations
did not mention noticeable NaI interstellar lines and 
$(B-V)$ color near maximum was quite blue: $\sim 0.1$ mag. 
Adopting distance
$\mu=32.07$ from LEDA, Galactic extinction $A_B=0.55$ from
Schlegel et al. (1998), we derive $M_V=-17.5$, which is 
close to the mean value for SN Ib/c (Richardson et al., 2002)

\medskip

{\bf SN 2001V}. The light curves, presented in Fig. 7, can be
matched closely by those for SN 1991T. Our data are in good
agreement with the results by Vinko et al. (2003), only in the $R$
band some systematic difference can be noticed. Our magnitudes for
comparison star 1 are in excellent agreement with data by Vinko et al.
(2003), with maximum difference of only 0.01 mag.    
The magnitudes at maximum light can be determined from template
curve fitting, as there was a gap in observations near maximum:
$B_{max}=14.7, V_{max}=14.6, R_{max}=14.7, I_{max}=14.9$.
The time of maximum in $B$ was about JD 2451974. SN 2001V was
clearly a slow declining SN Ia, but as the maximum was not
covered with observations, we would not attempt to derive 
$\Delta m_{15}(B)$, but estimate only $\beta_V=0.053$. This value
is similar to the one for SN 2000E. 
The extinction for SN 2001V is very small both in our Galaxy and
in the host galaxy: Galactic $A_B=0.085$ according to Schlegel et al. 
(1998), and the color curve $(B-V)$ does not show significant
reddening. Vinko et al. (2003) derived total reddening $E(B-V)=0.05$.
Adopting this value and distance modulus $\mu=34.11$ from LEDA,
we obtain $M_B=-19.6, M_V=-19.7$ -- the same values as for SN 2000E.
So photometric characteristics of these two SNe Ia are nearly
identical. 
 
\medskip

{\bf SN 2001X}. This SN was certainly a type II-P event, as can
be seen from Fig. 8, where we also plotted data obtained by
amateur astronomers in the bands close to $V$ and $R$ from 
{\tt http://www.astrosurf.com/snweb2/2001/01X\_/01X\_Meas.htm}.
Most of their magnitudes are in good agreement with our 
results. The light curves are fitted with those for 
typical SN II-P 1999gi (Leonard et al., 2002). We can estimate 
$B_{max}=15.2$ on JD 2451974, $V_{max}=15.2, R_{max}=14.9,
I_{max}=14.7$. The plateau phase lasted approximately until
JD 2452077, that is about 103 days, which is quite typical value
for type II-P SNe. 
The Galactic extinction in direction of SN 2001X amounts to 
$A_B=0.173$ according to 
Schlegel et al. (1998). Comparison of color $(B-V)$ curve for
SN 2001X with the one for SN 1999gi allows to conclude that
reddening of SN 2001X in the host galaxy was negligible.
Adopting $\mu=31.77$ from LEDA, we estimate $M_B=M_V=-16.7$,
close to the mean value for SN II-P (Richardson et al., 2002).

\medskip

The results of our study show that all investigated SNe 
are typical for their classes considering the shape of
their light curves and absolute magnitudes at maximum.

\medskip

{\bf Acknowledgements:}
This research has made use of the Lyon-Meudon Extragalactic Database
(LEDA) and of the Canadian Astronomy Data Centre. The author
is grateful to V.P.Goranskij, S.Yu.Shugarov and I.M.Volkov for
help in the observations.
The work was partly supported by RFBR grant 05-02-17480.  

\references

Burstein, D., Heiles, C., 1982, {\it Astron. J.}, {\bf 87}, 1165

Cappellaro, E., Turatto, M., Fernley, J., 1995, {\it IUE -- ULDA
Access Guide No. 6}, ESA Publications Division, ESTEC, 
Noordwijk, The Netherlands

Chevalier, C., Ilovaisky, S.A., 1991, {\it Astron.\& Astrophys. Suppl. Ser.},
{\bf 90}, 225

Chornock, R., Filippenko, A.V., 2001, {\it IAU Circ.}, No. 7577

Chornock, R., Li, W.D., Filippenko, A.V., 2001, {\it IAU Circ.}, No. 7593

Jha, S., Matheson, T., Challis, P., Kirshner, R., Berlind, P.,
2001, {\it IAU Circ.}, No. 7585

Landolt, A., 1992, {\it Astron. J.}, {\bf 97}, 337

Leonard, D.C., Filippenko, A.V., Li, W., et al., 2002, {\it Astron. J.},
{\bf 124}, 2490

Li, W., Fan, Y., Qiu, Y.L., Hu J.Y., 2001, {\it IAU Circ.}, No. 7591

Lira, P., Suntzeff, N.B., Phillips, M.M., et al., 1998, 
{\it Astron. J.}, {\bf 116}, 1006

Phillips, M.M., 2005, {\it ASP Conf. Ser.,} {\bf 342}, 211,
in 1604-2004: Supernovae as cosmological lighthouses, M.Turatto et al. eds.

Pskovskii, Yu.P., 1977, {\it Astron. Zhurn.}, {\bf 54}, 1188

Richardson, D., Branch, D., Casebeer, D., et al., 2002, {\it Astron. J.}
{\bf 123}, 745

Schlegel, D., Finkbeiner, D., Davis, M., 1998, {\it Astrophys. J.},
{\bf 500}, 525 

Stetson, P., 2000, {\it Publ. Astron. Soc. Pacif.}, {\bf 112}, 925

Turatto, M., Galletta, G., Cappellaro, E., 2000, {\it IAU Circ.},
No. 7351

Turatto, M., Benetti, S., Cappellaro, E., 2003, in Proc. of the
ESO/MPA/MPE Workshop, From Twilight to Highlight: The Physics of 
Supernovae, B.Leibundgut, W.Hillebrandt eds., 200 

Valentini, G., Di Carlo, E., Guidubaldi, D., et al., 2000, {\it IAU
Circ.}, No. 7351

Valentini, G., Di Carlo, E., Massi, F., et al., 2003, 
{\it Astrophys. J.}, {\bf 595}, 779 
 
Vinko, J., Csak, B., Csizmadia, S., et al., 2001, {\it Astron.\& Astrophys.},
{\bf 372}, 824

Vinko, J., Biro, I.B., Csak, B., et al., 2003, {\it Astron.\& Astrophys.},
{\bf 397}, 115

Xu, D.W., Qiu, Y.L., 2001, {\it IAU Circ.}, No. 7555 

\endreferences

\PZfig{8cm}{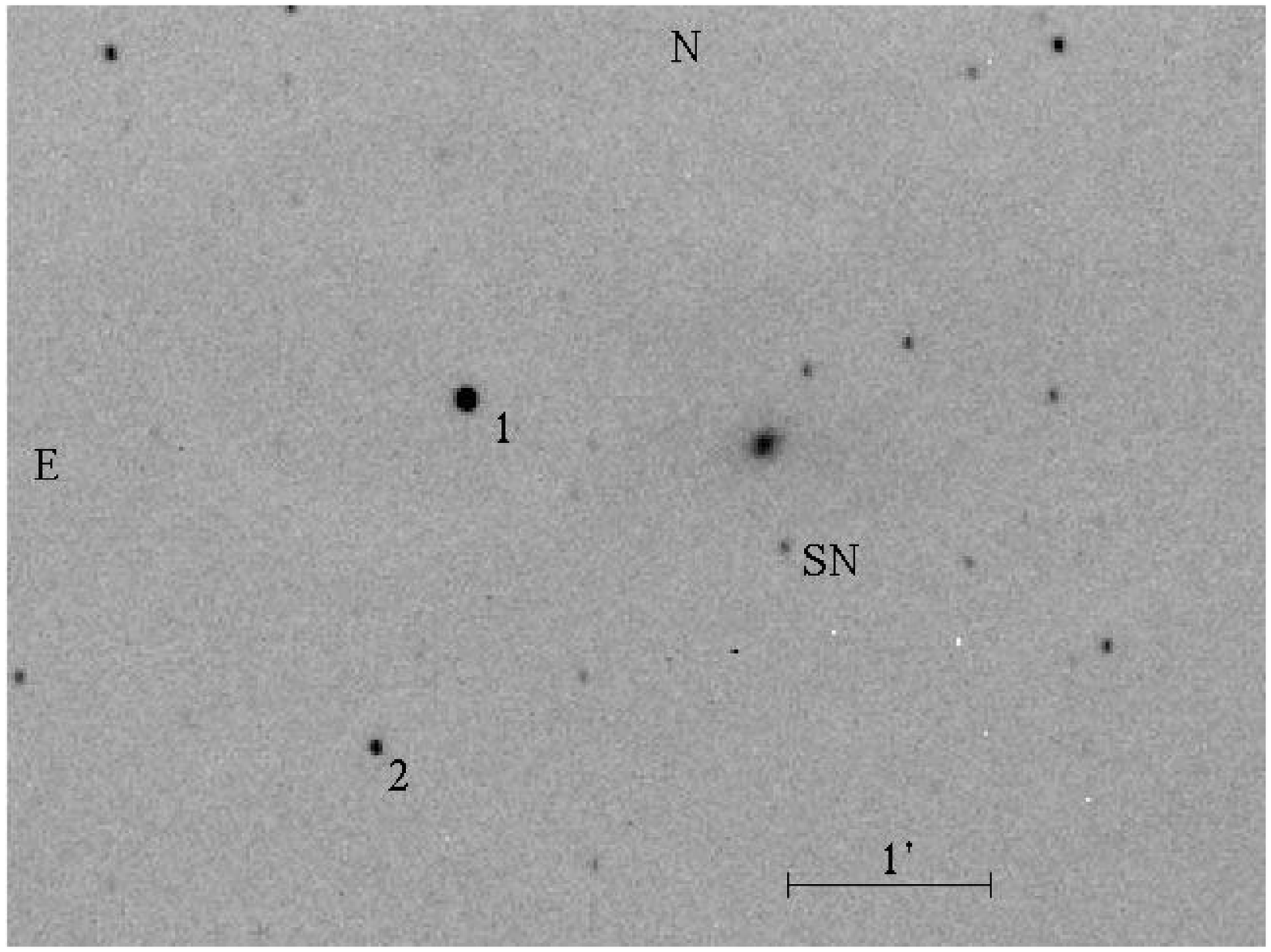}{SN 2000E in NGC 6951 with comparison
stars}

\PZfig{8cm}{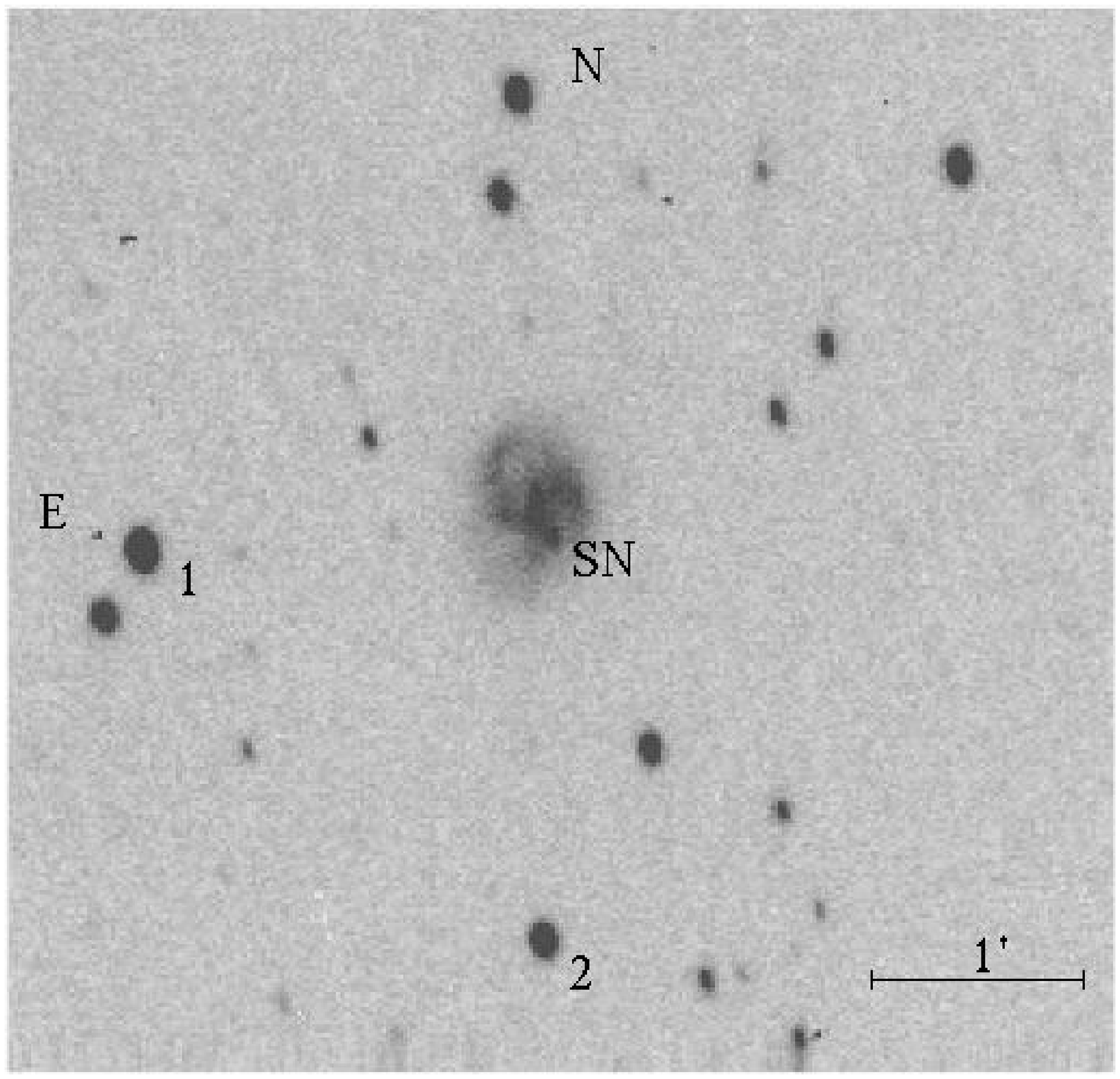}{SN 2001B in IC 391 with comparison
stars}

\PZfig{8cm}{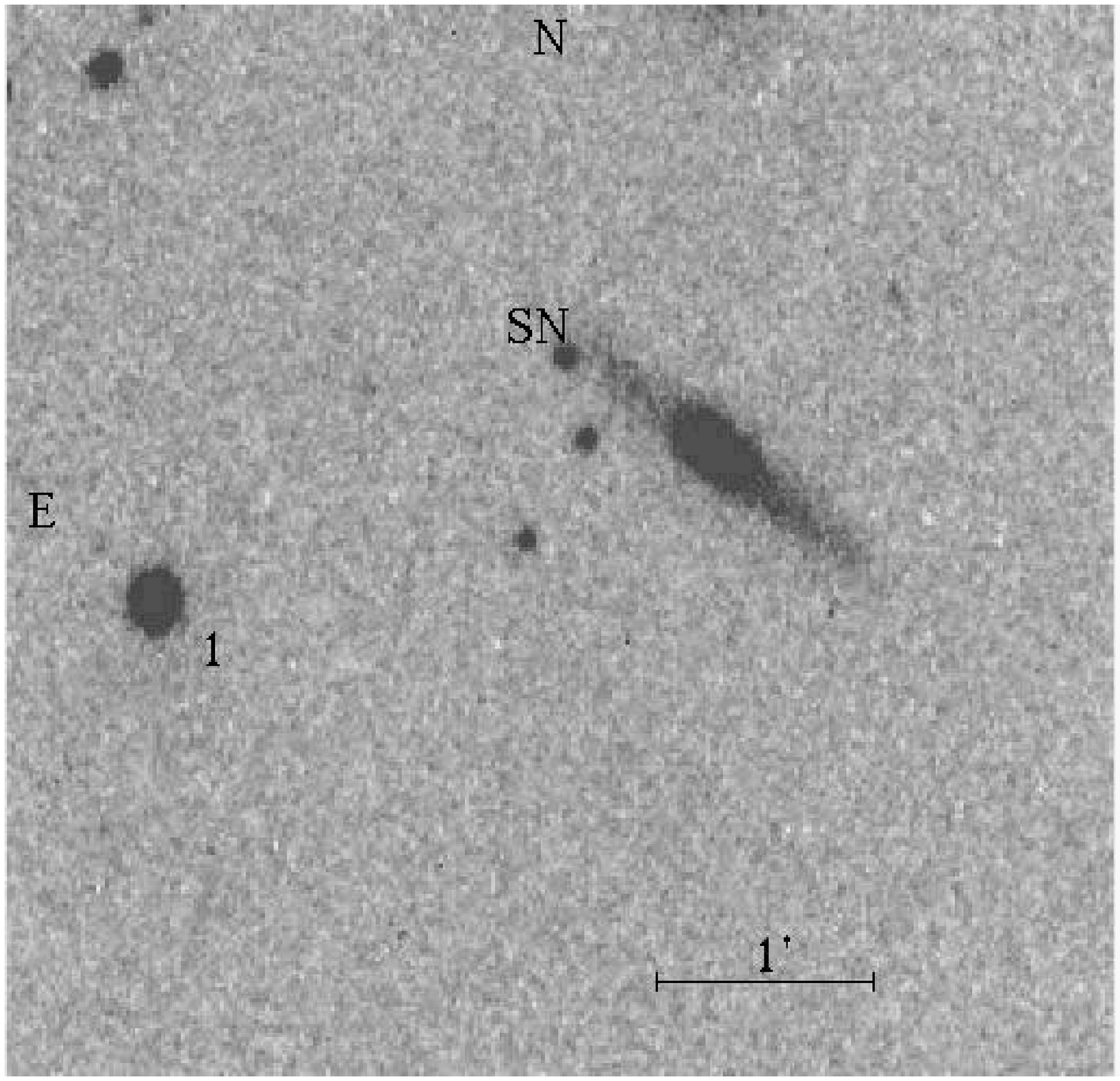}{SN 2001V in NGC 3987 with comparison
star}

\PZfig{8cm}{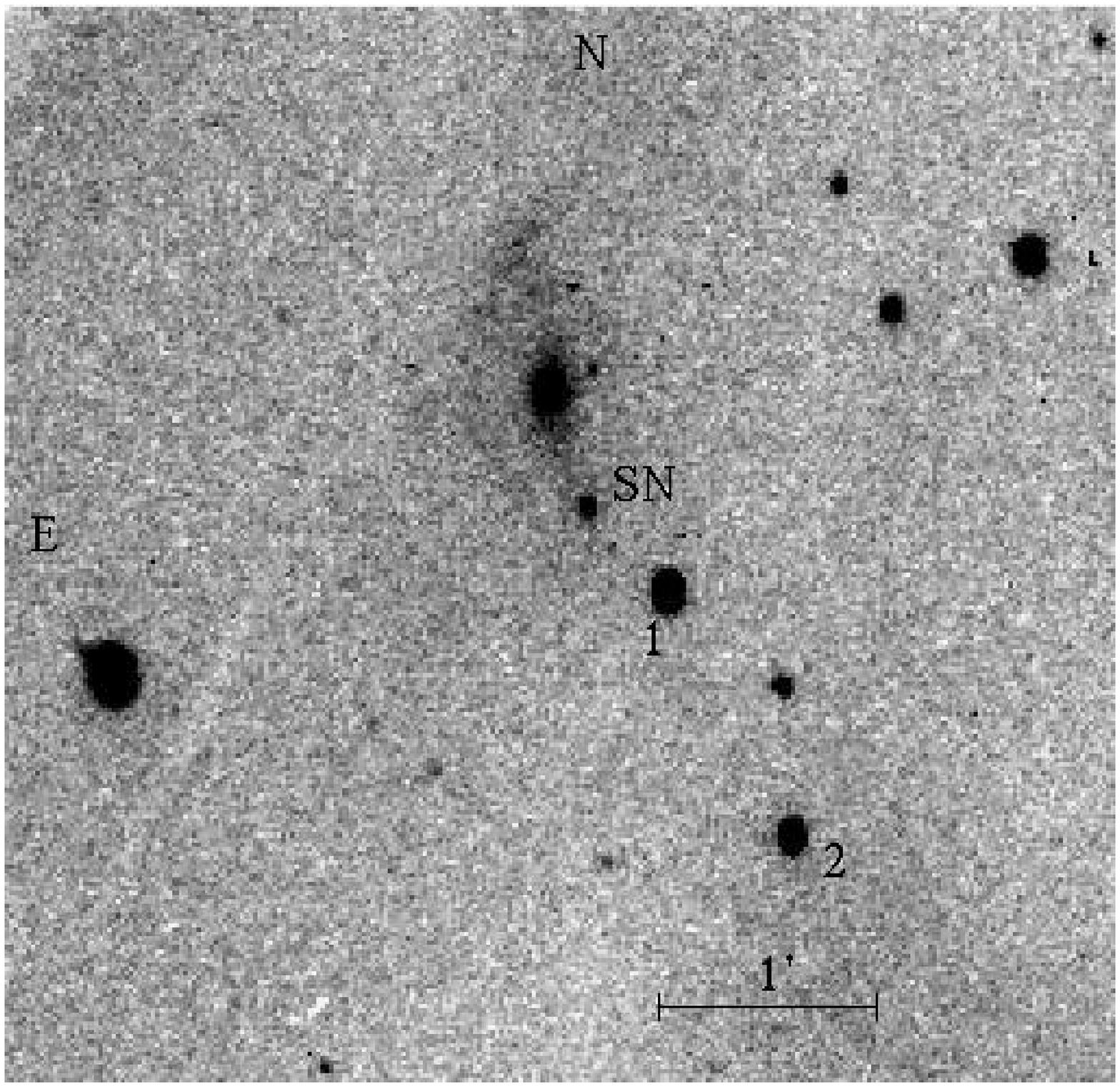}{SN 2001X in NGC 5921 with comparison
stars}

\PZfig{12cm}{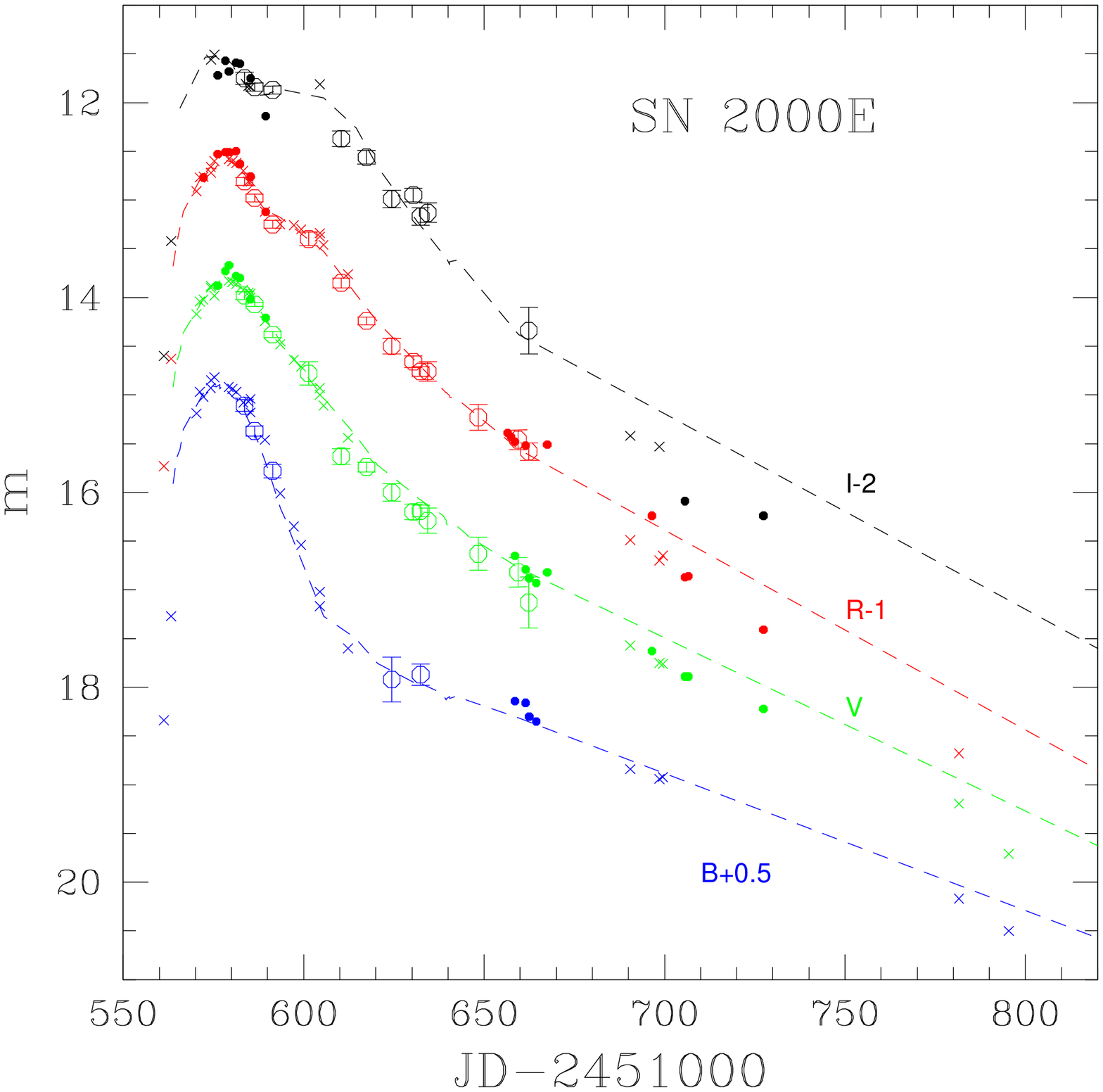}{$BVRI$ light curves of SN 2000E,
showing our photometry (circles) and that of
Vinko et al. (2001) (dots) 
and Valentini et al. (2003)(crosses). The dashed lines are the light curves of 
SN 1991T}

\PZfig{12cm}{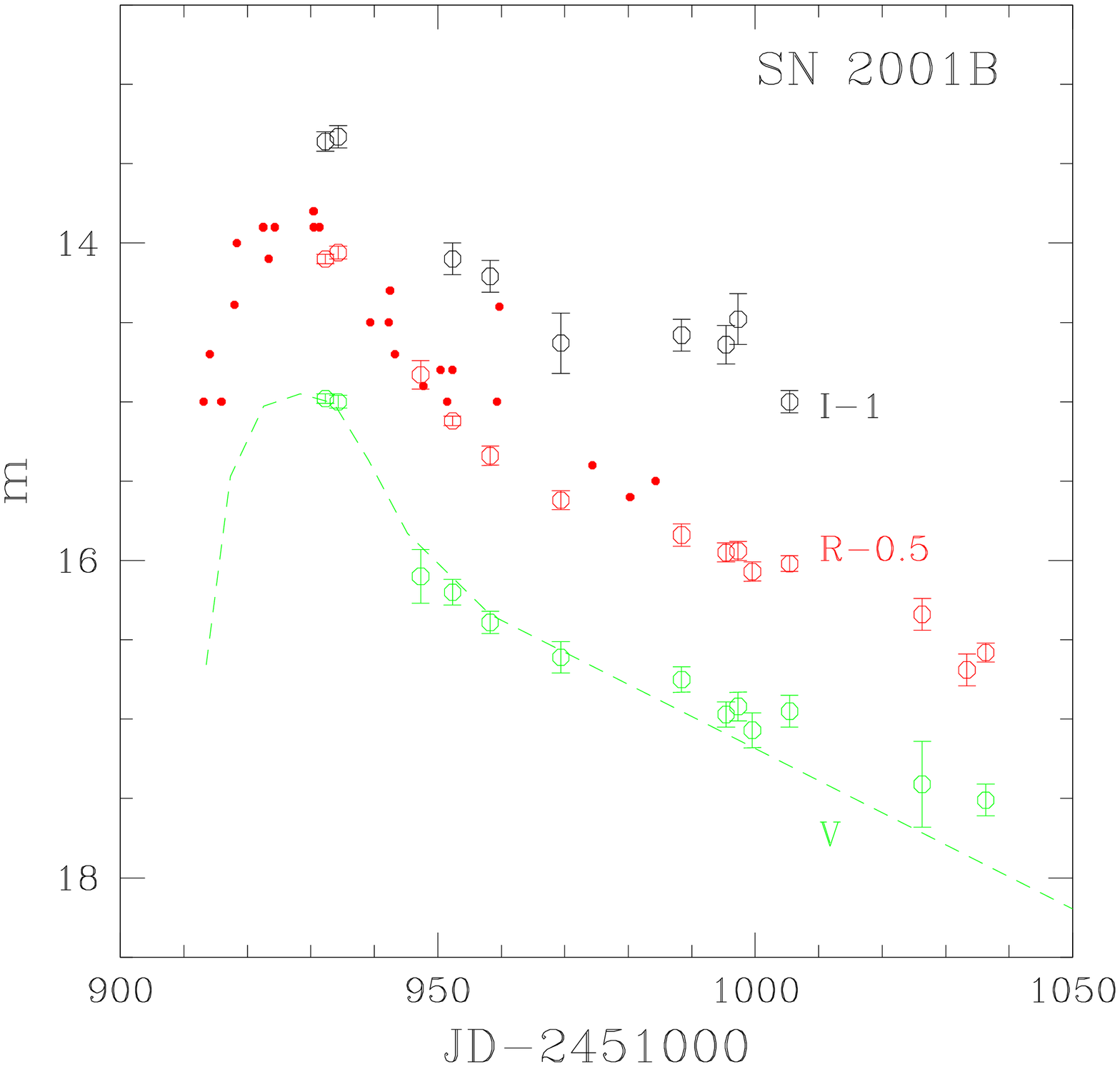}{$VRI$ light curves of SN 2001B. Circles show
our data, dots are for the observations of amateur astronomers.
The dashed line is $V$ light curve of SN 1983N}

\PZfig{12cm}{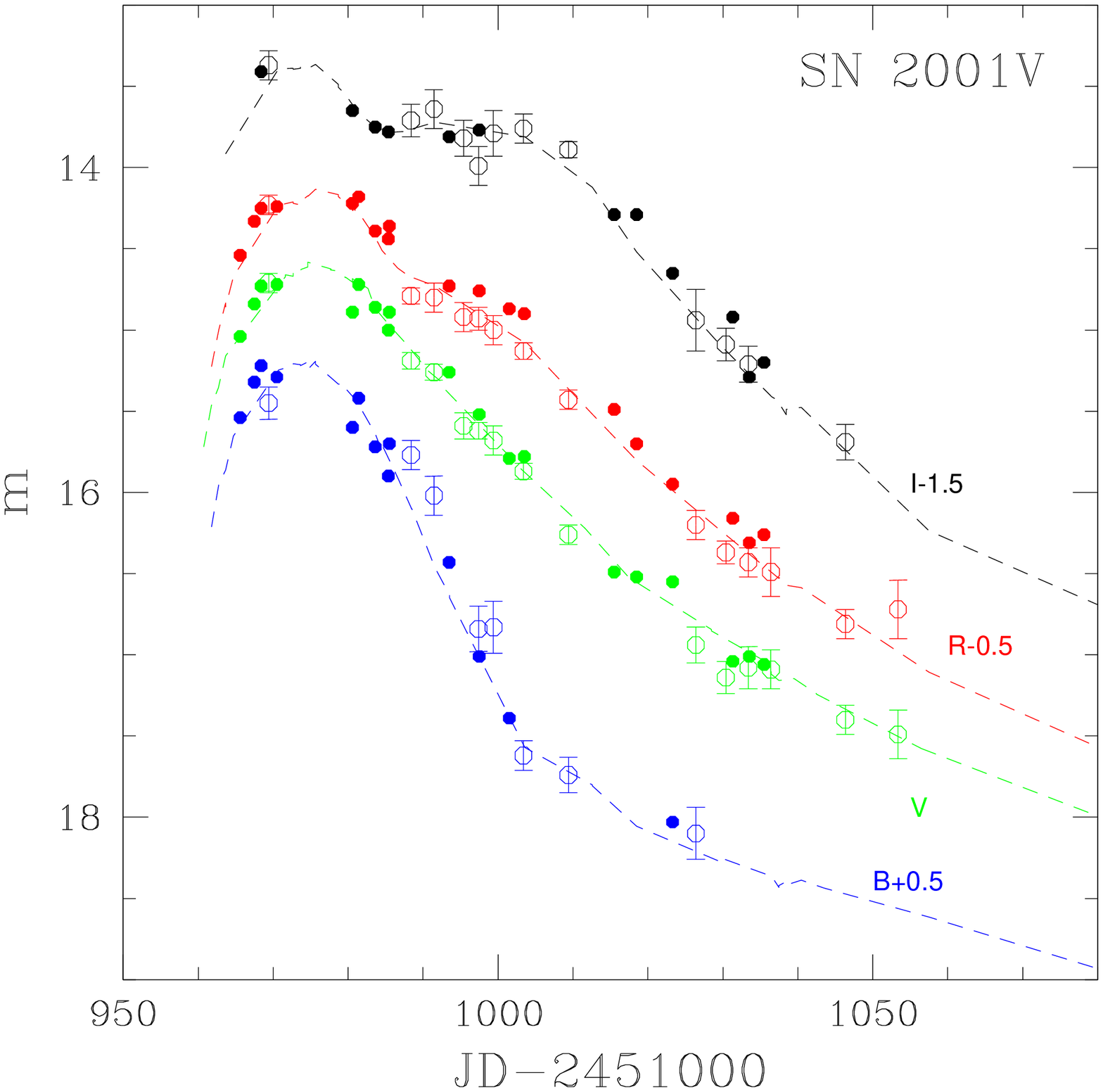}{$BVRI$ light curves of SN 2001V, 
showing our photometry (circles) and that of
Vinko et al. (2003)(dots).        
The dashed lines are the light curves of SN 1991T}          

\PZfig{12cm}{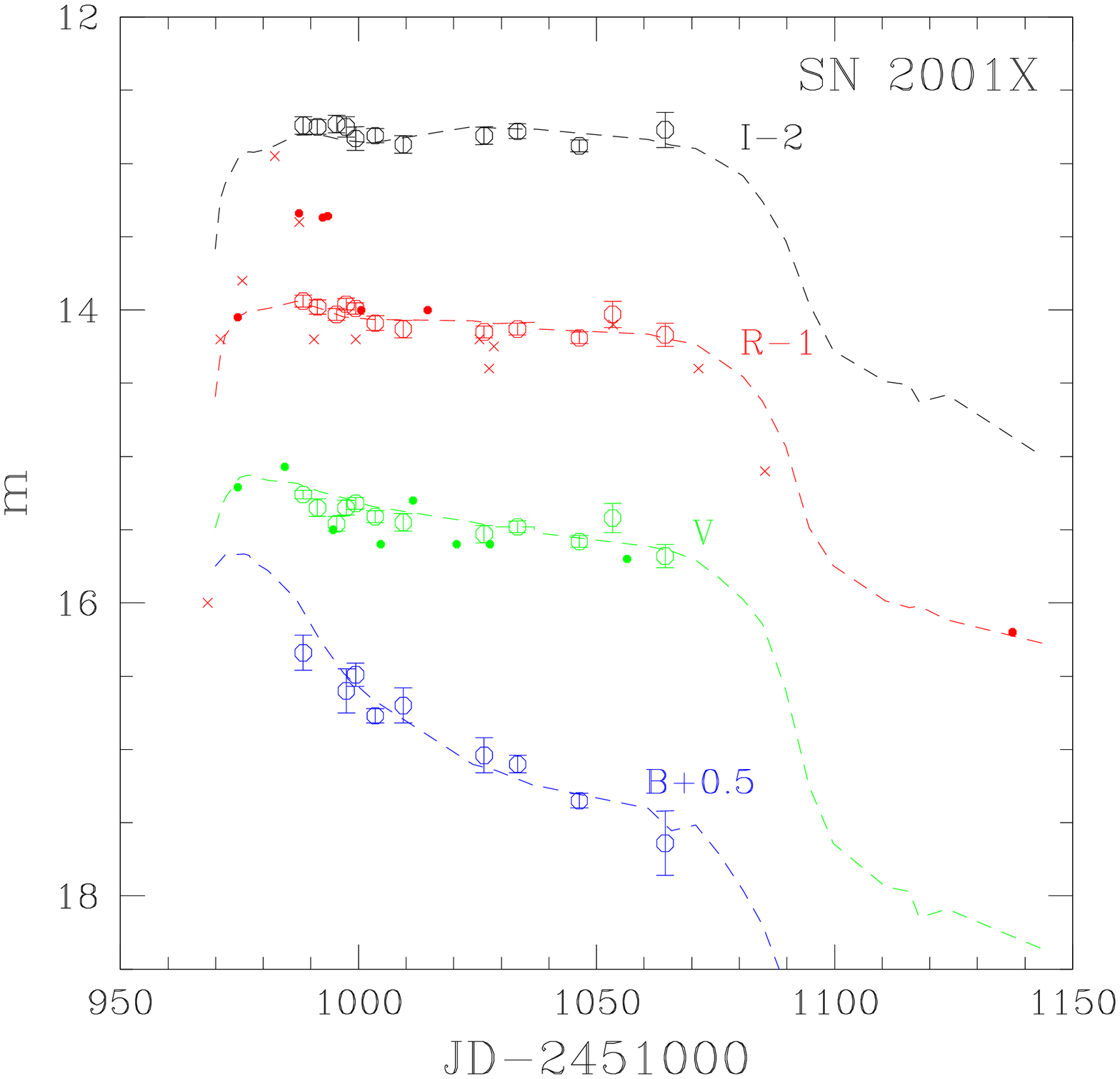}{$BVRI$ light curves of SN 2001X. Circles show
our data, dots are for observations of amateur astronomers in $V$ and
$R$ bands, crosses show photometry by amateurs with unfiltered CCDs.
The dashed lines are the light curves of SN 1999gi}

\newpage

\begin{table}
\centering
\caption{Color terms for reduction equations}\vskip2mm
\begin{tabular}{ccccc}
\hline
Tel., year & $K_b$ & $K_v$ & $K_R$ & $K_i$ \\
\hline
C60, 2000 & -0.53 & -0.045 & -0.50 & -0.53 \\
M30, 2000 & -0.32 & -0.046 & -0.56 & -0.06 \\
M30, 2001 & -0.31 & -0.002 & -0.42 & -0.23 \\
M70, 2001 & -0.11 & -0.008 & -0.34 & -0.39 \\
\hline
\end{tabular}
\end{table}

\begin{table}
\centering
\caption{Magnitudes of comparison stars}\vskip2mm
\begin{tabular}{lcccccccc}
\hline
Star & $B$ & $\sigma_B$ & $V$ & $\sigma_V$ & $R$ & $\sigma_R$ & $I$ & $\sigma_I$\\
\hline
NGC6951-1 & 14.35 & 0.02 & 12.61 & 0.02 & 11.67 & 0.02 & 10.86 & 0.03 \\
NGC6951-2 & 15.75 & 0.04 & 14.91 & 0.02 & 14.36 & 0.04 & 13.91 & 0.04 \\
IC391-1   & 14.01 & 0.08 & 13.24 & 0.02 & 12.85 & 0.01 & 12.50 & 0.04 \\
IC391-2   & 14.85 & 0.10 & 14.07 & 0.03 & 13.66 & 0.02 & 13.21 & 0.06 \\
NGC3987-1 & 13.08 & 0.05 & 12.23 & 0.01 & 11.74 & 0.01 & 11.27 & 0.03 \\
NGC5921-1 & 13.22 & 0.03 & 11.99 & 0.01 & 11.33 & 0.01 & 10.71 & 0.03 \\
NGC5921-2 & 13.51 & 0.04 & 12.90 & 0.02 & 12.57 & 0.02 & 12.24 & 0.05 \\
\hline
\end{tabular}
\end{table}

\begin{table}
\caption{Observations of SN 2000E}\vskip2mm
\begin{tabular}{ccccccccccc}
\hline
JD 2450000+ & $B$ & $\sigma_B$ & $V$ & $\sigma_V$ & $R$ & $\sigma_R$ & 
$I$ & $\sigma_I$ & Tel.\\
\hline
1583.62 &  14.61 & 0.06 &  13.98 & 0.04 &  13.81&  0.04 &  13.75 & 0.06 & C60  \\
1586.61 &  14.87 & 0.05 &  14.07 & 0.02 &  13.98&  0.04 &  13.84 & 0.04 & C60  \\
1591.60 &  15.28 & 0.07 &  14.38 & 0.03 &  14.25&  0.04 &  13.87 & 0.04 & C60  \\
1601.60 &        &      &  14.78 & 0.12 &  14.40&  0.07 &        &      & M30  \\
1610.60 &        &      &  15.63 & 0.08 &  14.85&  0.05 &  14.37 & 0.08 & M30  \\
1617.57 &        &      &  15.74 & 0.05 &  15.24&  0.04 &  14.56 & 0.07 & M30  \\
1624.57 &  17.42 & 0.23 &  16.00 & 0.09 &  15.50&  0.08 &  14.99 & 0.09 & M30  \\
1630.58 &        &      &  16.20 & 0.08 &  15.66&  0.06 &  14.95 & 0.07 & M30  \\
1632.53 &  17.37 & 0.11 &  16.19 & 0.06 &  15.76&  0.05 &  15.17 & 0.09 & M30  \\
1634.53 &        &      &  16.29 & 0.13 &  15.76&  0.10 &  15.13 & 0.10 & M30  \\
1648.39 &        &      &  16.63 & 0.17 &  16.23&  0.13 &        &      & M30  \\
1659.41 &        &      &  16.82 & 0.15 &  16.46&  0.10 &        &      & M30  \\
1662.40 &        &      &  17.13 & 0.26 &  16.58&  0.09 &  16.34 & 0.24 & M30  \\
\hline
\end{tabular}
\end{table}

\begin{table}
\caption{Observations of SN 2001B}\vskip2mm
\begin{tabular}{ccccccccccc}
\hline
JD 2450000+ & $B$ & $\sigma_B$ & $V$ & $\sigma_V$ & $R$ & $\sigma_R$ &
$I$ & $\sigma_I$ & Tel.\\
\hline
1932.29 &  15.27 & 0.15 &  14.98 &0.03  &  14.60& 0.03  &  14.36& 0.06 & M30 \\
1934.30 &  14.94 & 0.18 &  15.00 &0.04  &  14.56& 0.04  &  14.33& 0.07 & M30 \\
1947.30 &        &      &  16.10 &0.17  &  15.33& 0.09  &       &      & M30 \\
1952.32 &        &      &  16.20 &0.08  &  15.62& 0.03  &  15.10& 0.10 & M30 \\
1958.25 &        &      &  16.39 &0.07  &  15.84& 0.06  &  15.21& 0.10 & M30 \\
1969.39 &        &      &  16.61 &0.10  &  16.12& 0.06  &  15.63& 0.19 & M30 \\
1988.41 &        &      &  16.75 &0.08  &  16.34& 0.07  &  15.58& 0.10 & M30 \\
1995.40 &        &      &  16.97 &0.08  &  16.45& 0.06  &  15.64& 0.12 & M30 \\
1997.31 &        &      &  16.92 &0.09  &  16.44& 0.06  &  15.48& 0.16 & M30 \\
1999.55 &        &      &  17.07 &0.11  &  16.57& 0.06  &       &      & M30 \\
2005.41 &        &      &  16.95 &0.10  &  16.52& 0.05  &  16.00& 0.07 & M70 \\
2026.31 &        &      &  17.41 &0.27  &  16.84& 0.10  &       &      & M70 \\
2033.35 &        &      &        &      &  17.19& 0.10  &       &      & M70 \\
2036.32 &        &      &  17.51 &0.10  &  17.08& 0.06  &       &      & M70 \\
\hline
\end{tabular}
\end{table}

\begin{table}
\caption{Observations of SN 2001V}\vskip2mm
\begin{tabular}{ccccccccccc}
\hline
JD 2450000+ & $B$ & $\sigma_B$ & $V$ & $\sigma_V$ & $R$ & $\sigma_R$ &
$I$ & $\sigma_I$ & Tel.\\
\hline
1969.55 &  14.95 & 0.10 &  14.71 & 0.06 &  14.73 & 0.06 & 14.87 & 0.09 & M30 \\
1988.47 &  15.27 & 0.09 &  15.19 & 0.05 &  15.29 & 0.05 & 15.21 & 0.10 & M30 \\
1991.45 &  15.52 & 0.12 &  15.26 & 0.05 &  15.30 & 0.09 & 15.14 & 0.12 & M30 \\
1995.45 &        &      &  15.59 & 0.08 &  15.42 & 0.09 & 15.32 & 0.11 & M30 \\
1997.40 &  16.34 & 0.14 &  15.62 & 0.05 &  15.43 & 0.07 & 15.49 & 0.12 & M30 \\
1999.44 &  16.33 & 0.16 &  15.68 & 0.09 &  15.50 & 0.09 & 15.29 & 0.14 & M30 \\
2003.47 &  17.12 & 0.09 &  15.87 & 0.05 &  15.63 & 0.05 & 15.26 & 0.09 & M70 \\
2009.41 &  17.24 & 0.11 &  16.26 & 0.06 &  15.93 & 0.06 & 15.39 & 0.05 & M70 \\
2026.37 &  17.60 & 0.16 &  16.94 & 0.11 &  16.70 & 0.09 & 16.44 & 0.19 & M70 \\
2030.32 &        &      &  17.14 & 0.10 &  16.87 & 0.07 & 16.59 & 0.10 & M70 \\
2033.41 &        &      &  17.08 & 0.13 &  16.93 & 0.09 & 16.71 & 0.11 & M70 \\
2036.41 &        &      &  17.09 & 0.12 &  16.99 & 0.15 &       &      & M70 \\
2046.34 &        &      &  17.40 & 0.09 &  17.31 & 0.09 & 17.19 & 0.11 & M70 \\
2053.38 &        &      &  17.49 & 0.15 &  17.22 & 0.18 &       &      & M30 \\
\hline
\end{tabular}
\end{table}

\begin{table}
\caption{Observations of SN 2001X}\vskip2mm
\begin{tabular}{ccccccccccc}
\hline
JD 2450000+ & $B$ & $\sigma_B$ & $V$ & $\sigma_V$ & $R$ & $\sigma_R$ &
$I$ & $\sigma_I$ & Tel.\\
\hline
1988.57 & 15.84 & 0.12 &   15.26 & 0.03 &   14.94 & 0.04 & 14.74&  0.06& M30\\
1991.50 &       &      &   15.35 & 0.06 &   14.98 & 0.05 & 14.75&  0.06& M30\\
1995.51 &       &      &   15.46 & 0.05 &   15.03 & 0.04 & 14.73&  0.06& M30\\
1997.46 & 16.10 & 0.15 &   15.35 & 0.05 &   14.96 & 0.04 & 14.75&  0.07& M30\\
1999.50 & 15.99 & 0.08 &   15.32 & 0.04 &   14.99 & 0.04 & 14.83&  0.08& M30\\
2003.51 & 16.27 & 0.06 &   15.41 & 0.04 &   15.09 & 0.05 & 14.81&  0.05& M70\\
2009.53 & 16.20 & 0.12 &   15.45 & 0.06 &   15.13 & 0.06 & 14.87&  0.06& M70\\
2026.46 & 16.54 & 0.12 &   15.53 & 0.06 &   15.15 & 0.04 & 14.81&  0.06& M70\\
2033.46 & 16.60 & 0.07 &   15.48 & 0.04 &   15.13 & 0.04 & 14.78&  0.05& M70\\
2046.43 & 16.85 & 0.06 &   15.58 & 0.04 &   15.19 & 0.04 & 14.88&  0.05& M70\\
2053.42 &       &      &   15.42 & 0.10 &   15.03 & 0.09 &      &      & M30\\
2064.40 & 17.14 & 0.22 &   15.68 & 0.08 &   15.17 & 0.08 & 14.77&  0.12& M30\\
\hline
\end{tabular}
\end{table}
\end{document}